\documentclass[prd, preprint,
aps,
amsmath,
amssymb,
nofootinbib,
superscriptaddress,
]{revtex4-2}
\usepackage{graphicx,dcolumn,bm}
\usepackage{amsfonts,amsmath,amssymb,amsthm,amscd}
\usepackage[english]{babel}
\usepackage{physics}
\usepackage{needspace}
\usepackage{xcolor}
\usepackage{bm}
\usepackage{slashed}
\usepackage[export]{adjustbox}

\definecolor{linkblue}{HTML}{2e3092}
\usepackage[colorlinks=true, allcolors = linkblue]{hyperref}
\renewcommand{\vec}[1]{\bm{#1}}

\newcommand{\mJ}{\mathcal{J}}
\AtBeginDocument{\renewcommand{\natexlab}[1]{#1}}

\allowdisplaybreaks

\begin{document}
\title{Structure of radiative corrections in a strong constant crossed field in the bubble-chain approximation}
\author{A.~A. Mironov}\email{mironov.hep@gmail.com}
\affiliation{Prokhorov General Physics Institute of the Russian Academy of Sciences, Moscow, 119991, Russia}
\affiliation{National Research Nuclear University MEPhI, Moscow, 115409,  Russia}
\affiliation{Steklov Mathematical Institute, Russian Academy of Sciences, Gubkina str. 8, Moscow, 119991, Russia}
\author{A.~M.~Fedotov}
\affiliation{National Research Nuclear University MEPhI, Moscow, 115409,  Russia}

\begin{abstract}
We present a calculation of the electron mass operator combining the bubble-chain photon and the leading order electron propagators in a strong constant crossed electromagnetic field. The photon propagator is obtained by a summation of the Dyson series over 1PI polarization loop insertions and is accounted for in general form. By applying the derived mass operator, we find the bubble-chain electron propagator and the elastic scattering amplitude. The results are expressed in the general form of expansion over $\gamma$-matrix structures, each multiplied by a scalar invariant function of the electron virtuality and the quantum dynamical parameter $\chi$. We perform their asymptotic analysis and identify the dominant contribution at $\chi\gg 1$. The presented calculations can be generalized by replacing the leading order electron propagator with the bubble-chain one. This will allow a consistent study the Dyson-Schwinger equations in the fully nonperturbative regime $\alpha\chi^{2/3}\gtrsim1$ within the bubble-chain approximation. In addition, we accompany our paper with the open-access computer-algebraic scripts containing all the presented computations [\url{https://github.com/ArsenyMironov/SFQED-Loops}].
\end{abstract}

\maketitle

\section{Introduction}

QED is known for its tremendously precise predictions allowing for testing the Standard Model \cite{parker2018measurement}. At its core, QED relies on the perturbative expansion in the fine structure constant\footnote{We use units such that $\hbar=c=\varepsilon_0=1$, electron mass and charge are denoted by $m$ and $-e$ respectively ($e>0$), and the signature of the Minkowski metric is $(+,-,-,-)$.} $\alpha=e^2/4\pi\approx 1/137$, which breaks down only at the scale of the Landau pole \cite{gies2020asymptotically}. A strong electromagnetic background may severely impact the QED perturbative expansion \cite{di2012extremely}. For instance, in a plane wave type field of frequency $\omega$ and amplitude $E$, it breaks down as the dimensionless field strength $a_0=eE/m\omega$ becomes $\gtrsim 1$. This issue is cured by a proper resummation over all possible interactions with the background, namely, in the Furry picture \cite{furry1951bound}. Such an approach establishes the framework of strong-field QED (SFQED).

In effect, in SFQED one still expands in powers of $\alpha$ as in QED, while fermion lines are replaced with the ones dressed with the external field \cite{di2012extremely,narozhny2015extreme}. Once the (asymptotically stable) dressed fermion states are explicitly defined, it is possible to calculate amplitudes of various field-induced or modified scattering processes. In backgrounds, which can be (approximately) replaced by a constant crossed field (CCF), they are controlled by the quantum dynamical parameter  $\chi=e\sqrt{-(Fp)^2}/m^3$, where $p$ is the particle 4-momentum and  $F$ is the external field strength tensor. The parameter $\chi$ quantifies the particle rest frame field strength (normalized by the QED critical field $F_0= m^2/e$). At $\chi\ll 1$, the dominating effect is the first-order nonlinear photon emission by charged particles, which can be effectively treated as classical radiation \cite{nikishov1964quantum, blackburn2020radiation}, whereas at $\chi\gtrsim 1$, quantum recoil and the first-order nonlinear $e^-e^+$ pair photoproduction become equally significant \cite{zhang2020relativistic}.

At large values $\chi\gg1$, higher-order SFQED processes might also have an impact \cite{gonoskov2021charged}, as the one-loop photon polarization \cite{narozhny1969propagation} and electron mass  \cite{ritus1970mass} corrections both scale as $g=\alpha\chi^{2/3}$.
Such a strong scaling with $\chi$ is in sharp contrast to the logarithmic high-energy behaviour of field-free QED.
Further analysis have shown \cite{ritus1972radiative,ritus1972vacuum,narozhny1979radiation,narozhny1980expansion,mironov2020resummation} that $n$-loop radiative corrections scale as\footnote{The precise scaling is specific to the scattering amplitude under consideration. For the electron elastic scattering, it appears to be $g^{n} \chi^{-1/3}$ at $n\geq 3$ loop level \cite{mironov2020resummation}.} $g^n$ to all orders of perturbation theory. This striking
observation leads to important implications, also known as the Ritus-Narozhny (RN) conjecture \cite{fedotov2017conjecture}, that: (i) $g$ might be an effective expansion parameter of QED in a strong CCF; (ii) $g\gtrsim 1$ ($\chi\gtrsim 1600$) manifests a new fully nonpertupbative regime of radiation-matter interaction, in which radiative corrections become dominant and have to be resummed (see \cite{mironov2020resummation} for a deeper review).

According to the RN conjecture, the main contribution comes from the bubble-chain corrections, obtained by successive insertions of polarization loops to the photon lines \cite{narozhny1980expansion,mironov2020resummation}. Notably, such an insertion is gauge-invariant by construction. In our previous paper \cite{mironov2020resummation}, by considering the elastic electron scattering amplitude, we have shown that the bubble-chain corrections indeed scale as $g$ at $g<1$ to all orders of perturbation theory. It is still arguable whether other corrections should be taken into account too. For instance, the one-loop mass operator \cite{ritus1970mass} and vertex function \cite{morozov1981vertex, di2020one} in a CCF also scale as $g$ at $\chi\gg 1$. However, gauge invariance might lead to mutual cancellation of the asymptotically dominant terms in these corrections. Based on this argument, Narozhny conjectured that the vertex correction is asymptotically irrelevant  \cite{narozhny1979radiation,narozhny1980expansion}. Some evidence in favour of this statement was presented in the recent work \cite{di2020one}. In particular, it was shown by a direct calculation of the one-loop vertex correction in a CCF that the specific terms asymptotically growing as $g$ do not contribute to scattering amplitudes due to gauge invariance.

Physically, the replacement of a background by the CCF is justified by applying the locally constant field approximation. The type of dominating corrections might be inconsistent with the RN conjecture in the fields that do not fall within its scope  \cite{ilderton2020loop,edwards2021resummation,torgrimsson2021loops} or in theories deviating from standard QED \cite{ekman2020high}. Thus, high-$\chi$ behaviour of radiative corrections becomes qualitatively different in short or weak laser pulses \cite{podszus2019high,ilderton2019note}, reproducing QED-like logarithmic dependence. However, calculations in nonconstant fields beyond this approximation may be very challenging. For instance, the radiative corrections in a plane wave have been computed only at the one-loop level \cite{baier1976theory,baier1976interaction, meuren2013polarization, di2020one, di2021electron}. It is noteworthy, that even when the approximation is not applicable, a proper resummation of higher-order corrections still might be required in order to obtain a consistent result  \cite{heinzl2021classical,torgrimsson2021resummation1,torgrimsson2021resummation2}. Yet, the locally constant field approximation is robust in considerations of ultrarelativistic particles in strong backgrounds and valid as long as $a_0\gg \max (1,\chi^{1/3})$ \cite{ritus1985quantum}. Furthermore, it appears to be practical in many situations involving high-intensity optical lasers \cite{harvey2015testing, blackburn2018benchmarking, ilderton2019extended,di2021wkb}.

Experimentwise, while the state-of-the-art capabilities are at the level of $\chi\sim 1$ \cite{meuren2020seminal,abramowicz2021conceptual}, the regime $g\sim 1$ is considered to be within the reach of near future experiments using optical laser setups \cite{blackburn2019reaching,baumann2019laser,baumann2019probing}, at future lepton colliders \cite{yakimenko2019prospect}, or in the passing of high-energy electrons through aligned crystals \cite{di2019testing}. However, theoretical studies of the nonperturbative regime are limited. The standard in-out approach to calculation of scattering amplitudes is unreliable at $g\gtrsim1$, since the question about the stability of the asymptotic particle states becomes ambiguous. One may account for state damping by radiative corrections \cite{podszus2021first}, however, this approach is justified only in the narrow width approximation, namely, at $\chi\sim 1$. A rigorous consideration at $g\gtrsim1$ might involve some nonperturbative \cite{dunne2021higher}, instanton \cite{ahmadiniaz2020worldline} and nonequilibrium QFT \cite{fauth2021collisional} techniques, or a direct summation of, at least, the dominating corrections to loop diagrams \cite{mironov2020resummation} in combination with appropriately defined cutting rules that do not violate unitarity \cite{denner2015complex}.

\begin{figure}
	\begin{center}
		\includegraphics[width=0.9\linewidth]{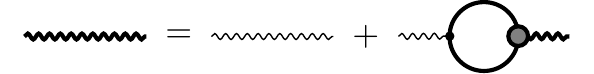}\\
		\includegraphics[width=0.9\linewidth]{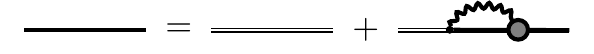}
	\end{center}
	\caption{The Dyson-Schwinger equations. The thin wavy and the double fermion lines corresponds to the leading order photon and (dressed by a CCF) electron propagators. The thick lines describe the exact photon and electron propagators. The filled circle corresponds to the full vertex. In the bubble-chain approximation it is replaced by the bare vertex (in a fixed gauge).}
	\label{fig:DS}
\end{figure}

The advantage of the bubble-chain approximation is that it allows, in principle, a consistent summation of the Dyson series for the photon and electron propagators, i.e. enclosing the Dyson-Schwinger (DS) equations as shown in Fig.~\ref{fig:DS}. In practice, to do so, we have to accomplish the following program: (i) determine the general $\gamma$-matrix and tensor structure of the exact photon and electron propagators; (ii) impose a gauge-fixing condition in which the full vertex can be approximated by the bare one; (iii) calculate the electron mass and photon polarization operators with accounting for these propagators; (iv) plug the result into the DS equations and reformulate them by exploiting the specific structure of the constituents. If the resulting equations are solvable, exactly or approximately, then hopefully this will shed some light on phenomena arising in the fully nonperturbative regime. Though not directly relevant, let us mention that the DS approach appears to be successful in QED in a supercritical magnetic field \cite{gusynin1999dynamical, gusynin1999theory}.

In this paper, we take a step towards the bubble-chain DS equations by (partially) considering the points (i) and (iii). In particular, we introduce the dressed 1PI bubble-chain photon propagator of a general form. We plug this propagator into the mass operator in a CCF with accounting the leading-order (LO) electron propagator and compute it. Then we present the corresponding bubble-chain electron propagator and analyse its implications to the DS equations. While in our previous work \cite{mironov2020resummation} we studied the asymptotic properties the electron elastic scattering amplitude obtained by a resummation of the one-loop polarization insertions, here we present a more general off-shell calculation. Furthermore, we adopt different evaluation sequence that appears to be more suited for multi-loop calculations in the bubble-chain approximation. 

Let us also note that the complexity of calculations in SFQED rapidly grows with the order of perturbation theory. Even in the simplest case of a CCF, intermediate computations appear to be too lengthy to be presented in print. Motivated by this, we developed several computer algebraic scripts with the aid of the FeynCalc package \cite{mertig1991feyn,shtabovenko2016new,shtabovenko2020feyncalc}. They contain the full version of the calculations discussed here. The scripts are open-access \cite{github}. We believe that the presented results and the developed scripts make a solid basis for advancing further the studies of the bubble-chain DS equations in a CCF.

The paper is organized as follows. We start with
introducing the notations in Sec.~\ref{sec:notations}. In Sec.~\ref{sec:e_propagator} we present the LO electron propagator in a CCF in the proper time representation. In the follow-up Sec.~\ref{sec:photon_propagator} we discuss the bubble-chain photon propagator obtained by resummation of 1PI polarization loop insertions and derive it in the proper time representation. We apply these propagators in Sec.~\ref{sec:mass_op} in order to calculate the bubble-chain mass operator. Then in Sec.~\ref{sec:exact_electron_propagator} we present the bubble-chain electron propagator and discuss its general structure. After that we derive the electron bubble-chain elastic scattering amplitude in Sec.~\ref{sec:amplitude}. In Sec.~\ref{sec:discussion} we perform an asymptotic analysis of the obtained results and identify the terms dominating at $\chi\gg 1$, and also give some outlook for further advancing towards a consistent study of the DS equations. We conclude our paper in Sec.~\ref{sec:conclusions}. We relegate the derivation and the explicit answer for the bubble-chain electron propagator in the proper time representation to Appendix~\ref{app1}.

\section{Electron and photon propagators}
\label{sec:propagators}
\subsection{Notations}
\label{sec:notations}
Let us consider an electron propagating in a CCF. We choose a fixed gauge for the external field, such that a 4-potential is given by $A^\mu(\varphi)=a^\mu \varphi$. Here, $a^\mu$ is a constant 4-vector and the phase $\varphi=kx$, where $k^\mu$ is a lightlike 4-vector satisfying $k^2=ka=0$. The corresponding field strength tensor reads $F^{\mu\nu}=k^\mu a^\nu-k^\nu a^\mu$. Without loss of generality, we may introduce a special reference frame (RF) where $k^\mu=m(1,0,0,1)$ and $a^\mu=(0,-\vec{a}_\perp,\,0)$ with $\vec{a}_\perp=(a^1,\,a^2)$. In this RF, for a 4-vector $p^\mu$ we have scalar products $(kp)= m p_-$ and $(ap)=-\vec{a}_\perp\vec{p}_\perp$, where $p_-=p^0-p^3$ and $\vec{p}_\perp=(p^1,\,p^2)$. Also, for an arbitrary $q^\mu$ we have $(pq)=p_+q_-+p_-q_+-\vec{p}_\perp\vec{q}_\perp$, where $p_+=(p^0+p^3)/2$ (the same notations go for $q$). For any 4-vector $p$, we will refer to $p_{\pm,\perp}$ as the light-cone variables (similarly to e.g.  Ref.~\cite{di2020one}).

In our calculations, we use dimensional regularization of divergent integrals \cite{veltman1972regularization, peskin2018introduction}. By $D=4-\varepsilon$ we denote the fractional dimension, implying the limit $\varepsilon\rightarrow0$ at the end of calculation. Therefore, the Minkowski metric satisfies $g^\mu_{\hphantom{\mu}\mu}=D$, while all the scalar products are written as usual, e.g. $p^\mu p_\mu=p^2$. We define $D$-dimensional gamma matrices $\gamma^\mu$, so that their anticommutator $\lbrace \gamma^\mu,\gamma^\nu\rbrace = 2g^{\mu\nu}$, and the trace $\mathrm{Tr}\,\gamma^\mu\gamma^\nu=4g^{\mu\nu}$. As a consequence, the gamma matrices obey the identities  $\gamma_\mu\gamma^\mu=D$, $\gamma_\mu\gamma^\rho \gamma^\mu=(D-2)\gamma^\rho$ etc. The $\gamma^5=i\gamma^0\gamma^1\gamma^2\gamma^3$ matrix, which is also present in our calculations, might need a special treatment in $D$-dimensions \cite{breitenlohner1977dimensional}. However, as one will see further, $\gamma^5$ will arise only in a product with the Levi-Civita tensor $\varepsilon^{\mu\nu\delta\lambda}$ ($\varepsilon^{0123}=1$), therefore causing no additional difficulties. Moreover, for simplicity, we will treat it as a $D=4$ dimensional object since it will enter only regular terms.

\subsection{Electron propagator in a CCF} 
\label{sec:e_propagator}
The LO propagator of an electron $S_0^c(x'',x')$ in an external field obeys the equation
\begin{equation}
\label{S_eq}
\left[(\gamma\hat{p})+e(\gamma A)-m\right]S_0^c(x'',x')=i\delta(x''-x'),
\end{equation}
where $\hat{p}_\mu=i\partial/\partial x^{\prime\prime\mu}$. A solution of this equation in a plane-wave background can be expressed in the Ritus $E_p$-representation \cite{ritus1972radiative} as
\begin{equation}
\label{Sc_Ep}
S_0^c(x'',x')=i\Lambda^{4-D} \int \frac{d^Dp}{(2\pi)^D}\frac{E_p(x'')\left[(\gamma p)+m\right]\bar{E}_p(x')}{p^2-m^2+i0},
\end{equation}
where we extended the definition of $S_0^c$ to $D=4-\varepsilon$ dimensions and introduced a mass scale $\Lambda$. The factor $\Lambda^{4-D}$ ensures that the natural dimension of $S_0^c$ is independent of $D$.

The matrix functions $E_p(x)$ multiplied by a free-electron bispinor $u_{p,\lambda}$ give the well-known Volkov solutions of the Dirac equation in a plane-wave \cite{volkov1935class}. In particular, we use  the $E_p$-functions in a CCF:
\begin{equation}
\label{Ep}
E_p(x'')=\left[1-\frac{e(\gamma k)(\gamma a)}{(kp)}\varphi'' \right]\exp\left\lbrace-ipx''+i\frac{e(ap)}{2(kp)}\varphi^{\prime\prime\,2}+i\frac{e^2a^2}{6(kp)}\varphi^{\prime\prime\,3}\right\rbrace,
\end{equation}
where $\varphi''=(kx'')$, and $\bar{E}_p(x')=\gamma^0 E^\dagger_p(x')\gamma^0$ is the Dirac conjugated function. 

Evaluation of the integrals in Eq.~\eqref{Sc_Ep} casts the propagator into the proper time representation \cite{schwinger1951gauge, ritus1972radiative}:
\begin{widetext}
\begin{equation}
\label{Sc}
\begin{split}
S_0^c(x'',x')=&e^{i(ax)\Phi} e^{-i \frac{\pi}{2}\frac{D-2}{2}} \frac{\Lambda^{4-D}}{(4\pi)^{D/2}}\int_0^\infty\frac{ds}{s^{D/2}}
\exp\left\lbrace-im^2 s-i\frac{x^2}{4s}+i\frac{s}{12}e^2\left(Fx\right)^2\right\rbrace
\\
&\times\left[m+\frac{(\gamma x)}{2s}-\frac{s}{3}e^2(\gamma F^2 x)+\frac{i}{2}mse(\sigma F)+\frac{i}{2}e(\gamma F^\star x)\gamma^5\right].
\end{split}
\end{equation}
\end{widetext}
Hereinafter, we adopt the notation $x=x''-x'$, $X=(x''+x')/2$, $\Phi=(kX)$.  $F^\star_{\mu\nu}=(1/2)\varepsilon_{\mu\nu\lambda\sigma}F^{\lambda\sigma}$ is the dual field strength tensor. For convenience, we introduced a shorthand notation for scalar combinations like $(\gamma F^2 x)=\gamma^\mu F_{\mu\nu}F^{\nu\delta}x_\delta$, and $(\sigma F)$ is the contraction of the field tensor $F_{\mu\nu}$ with the gamma matrix commutator $\sigma^{\mu\nu}=(i/2)(\gamma^\mu\gamma^\nu-\gamma^\nu\gamma^\mu)$.
Note that the electron proper time is defined by
\begin{equation}
\label{s}
s=\frac{x_-}{2p_-}=\frac{(kx)}{2(kp)}.
\end{equation}
Although the derivation of Eq.~\eqref{Sc} is straightforward, the detailed calculation can be found in \cite{github}.

\subsection{Bubble-chain photon propagator}
\label{sec:photon_propagator}
\begin{figure*}[tbp]
	\centering\includegraphics[width=\textwidth]{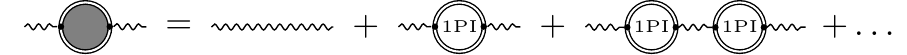}
	\caption{\label{fig:pol} The bubble-chain photon propagator obtained by a summation of 1PI polarization loop insertions in a CCF, see Eq.~\eqref{photon_prop}. }
\end{figure*}

The photon propagator $D^c$ with account for  1PI polarization corrections in a CCF obeys the Dyson-Schwinger equation (see Fig.~\ref{fig:pol}) \cite{ritus1972radiative}:
\begin{equation}
\label{Dyson_ph}
\left[l^2g^{\mu\nu}-\left(1-\frac1 d_l\right)l^\mu l^\nu-\Pi^{\mu\nu}(l)\right]D^c_{\nu\lambda}(l)=-i\delta_\lambda^\mu,
\end{equation}
where $l^\mu$ is the photon momentum, $d_l$ is the gauge-fixing parameter, and $\Pi^{\mu\nu}(l)$ is the 1PI photon polarization operator. In a CCF, $\Pi^{\mu\nu}(l)$ can be decomposed into the three transverse tensors:
\begin{equation}
\label{pi_expansion}
\begin{split}
\Pi_{\mu\nu}(l) &= \widehat{\Pi}(l^2,\chi_l) \left(l^2 g_{\mu\nu}-l_\mu l_\nu\right)+\sum\limits_{i=1}^2\Pi_i(l^2,\chi_l)\epsilon_\mu^{(i)}(l)\epsilon_\nu^{(i)}(l).
\end{split}
\end{equation}
The eigenfunctions $\widehat{\Pi}$ and $\Pi_{1,2}$ specifically depend on the virtuality $l^2$ and the dynamical quantum parameter $\chi_l=e\sqrt{-(F_{\mu\nu}l^\nu)^2}/m^{3}$ of the virtual photon, and the eigenvectors $\epsilon_\mu^{(1)}(l)=eF_{\mu\nu}l^\nu/(m^3\chi_l)$ and $\epsilon_\mu^{(2)}(l)=eF^\star_{\mu\nu}l^\nu/(m^3\chi_l)$ obey $\epsilon^{(i)\,2}=-1$, $(\epsilon^{(1)}\epsilon^{(2)})=0$. Notably, the second term in Eq.~\eqref{pi_expansion} is field-induced and should vanish at $F_{\mu\nu}\rightarrow0$.

By substituting the expansion \eqref{pi_expansion} into Eq.~\eqref{Dyson_ph} one obtains the 1PI bubble-chain (dressed) photon propagator in the momentum representation \cite{narozhny1969propagation, ritus1972radiative, shabad1975photon, mironov2020resummation}:
\begin{equation}
\label{photon_prop}
D^c_{\mu\nu}(l)= D_0(l^2,\chi_l) \left[g_{\mu\nu}-(1-d_l) \frac{l_\mu l_\nu}{l^2}\right]+\sum\limits_{i=1}^2 D_i(l^2,\chi_l)\epsilon_\mu^{(i)}(l)\epsilon_\nu^{(i)}(l).
\end{equation}
The renormalized photon propagator is given by \cite{mironov2020resummation}:
\begin{eqnarray}
\label{D_0}
D_0(l^2,\chi_l)=\frac{-i}{l^2+i0},\\
\label{D_i}
D_{1,2}(l^2,\chi_l)=\frac{i \Pi_{1,2}(l^2,\chi_l)}{\left(l^2+i0\right)\left[l^2-\Pi_{1,2}(l^2,\chi_l)\right]}.
\end{eqnarray}
Here, the argument $\chi_l$ is introduced to $D_0$ for uniformity of notations. In what follows, we omit the $l_\mu l_\nu$ term in the propagator \eqref{photon_prop} as it does not contribute to the resulting mass operator.

It is convenient to introduce  the dressed photon propagator in the $x$-representation:
\begin{equation}
\label{Dx_def}
D^c_{\mu\nu}(x)=\frac{\Lambda^{4-D}}{(2\pi)^D} \int d^D l\, D^c_{\mu\nu}(l)e^{-ilx}.
\end{equation}
Let us substitute Eq.~\eqref{photon_prop} into this expression and pass to the light-cone variables $l_{\pm,\perp}$, so that $d^Dl=(1/2|t|)dt\,dl^2\,d^{D-2}l_\perp$, where the photon proper time is defined by
\begin{eqnarray}
\label{t}
t=\frac{x_-}{2l_-}=\frac{\varphi}{2(kl)}.
\end{eqnarray}
 Note that the dynamical parameter of a photon in a CCF now reads
\begin{equation}
\label{chi_l}
\chi_l=\frac{\xi(kl)}{m^2}= \frac{\xi\varphi}{2 m^2 t},
\end{equation}
where $\xi^2=-e^2a^2/m^2$ is the dimensionless field strength parameter. The $D_{1,2}$ terms [see Eq.~\eqref{photon_prop}] that are quadratic in $l$ can be rewritten with the aid of differentiation:
\begin{equation}
\label{x_diff}
\int d^D l\,l_\alpha l_\beta D_i(l)e^{-ilx}=-\partial_\alpha \partial_\beta\int d^D l\, D_i(l)e^{-ilx}.
\end{equation}
Then, after expanding \[(lx)=l_-x_+ + l_+x_-  - \vec{l}_\perp \vec{x}_\perp=s(l^2+l_\perp^2)+(1/4t)(x^2+x_\perp^2)-\vec{l}_\perp \vec{x}_\perp\] in the exponent of Eq.~\eqref{Dx_def}, we can carry out the $(D-2)$-dimensional Gaussian integral over $\vec{l}_\perp$ straightforwardly.

Let us introduce dimensionless functions [cf. Ref.~\cite{mironov2020resummation}, Eq.~(31)]
\begin{eqnarray}
\label{J_def}
\mJ_n(t,\chi_l)= -i \int_{-\infty}^\infty dl^2\, D_n(l^2,\chi_l)e^{-il^2 t},\quad n=0,1,2,\\
\label{Jt_def}
\tilde{\mJ}_{1,2}(t,\chi_l)= \frac{i\left[\mJ_n(t,\chi_l)\right]'_t}{m^2}= -i \int_{-\infty}^\infty dl^2\, \frac{l^2}{m^2} D_{1,2}(l^2,\chi_l)e^{-il^2 t},
\end{eqnarray}
By using the functions \eqref{J_def} and performing differentiation over $x$ [see Eq.~\eqref{x_diff}], we finally arrive at the expression
\begin{widetext}
\begin{equation}
\label{Dx}
\begin{split}
D^c_{\mu\nu}(x)=  e^{-i \frac{\pi}{2}\frac{D-4}{2}} \frac{\Lambda^{4-D}}{(4\pi)^{D/2+1}}\int_0^\infty & \frac{dt}{t^{D/2}}  \exp\left(-i\frac{x^2}{4t}\right) \left\lbrace\mathcal{J}_0(t,\chi_l) g_{\mu\nu}  \vphantom{\frac{ (F^\star x)_\mu(F^\star x)_\nu}{m^2\xi^2 \varphi}}\right.\\
&  + \frac{\mathcal{J}_1(t,\chi_l)}{m^2\xi^2\varphi^2}e^2\left[(Fx)_\mu(Fx)_\nu -2it(F^2)_{\mu\nu}\right]\\
& \left. +  \frac{\mathcal{J}_2(t,\chi_l)}{m^2\xi^2\varphi^2}e^2\left[(F^\star x)_\mu(F^\star x)_\nu -2it(F^2)_{\mu\nu}\right]\right\rbrace.
\end{split}
\end{equation}
\end{widetext}
We will employ this result in the calculation of the resummed bubble-chain mass operator. A detailed derivation of Eq.~\eqref{Dx} is presented in \cite{github}. 

The functions $\mJ_n$ and $\tilde\mJ_{1,2}$ vanish at $t<0$, which follows from Eqs.~\eqref{D_0},~\eqref{D_i} and the fact that all the poles of $D_{1,2}$ are physical (an infinite number of them) \cite{ritus1972radiative}. In particular, $\mathcal{J}_0(t)=2\pi i\theta(\mathrm{Re}\,t)$. This reflects the photon propagator causality. At $\chi_l\gtrsim 1$ these functions can be approximated by the contribution from the main pole $l^2\approx \Pi_{1,2}(0,\,\chi_l)$ \cite{mironov2020resummation}:
\begin{eqnarray}
\label{J_res}
\mJ_{1,2}(t,\chi_l)\approx -2\pi i \theta\left(\mathrm{Re}\,t-t_{\mathrm{eff}}\right)\left[ e^{ -i \Pi_{1,2}(0,\chi_l)t} -1 \right],\\
\label{Jt_res}
\tilde{\mJ}_{1,2}(t,\chi_l)\approx -2\pi i \theta\left(\mathrm{Re}\,t-t_{\mathrm{eff}}\right)  \frac{\Pi_{1,2}(0,\chi_l)}{m^2} e^{ -i \Pi_{1,2}(0,\chi_l)t},
\end{eqnarray}
where the shift in the $\theta$-function argument roughly estimates the smearing of the causal $\theta$-function at the scale $t_{\mathrm{eff}}\sim 1/m^2\chi_l^{2/3}$ due to radiative corrections. 

\section{The bubble-chain radiative corrections}
\subsection{Mass operator}
\label{sec:mass_op}
\begin{figure*}[tbp]
	\centering\includegraphics[width=.4\textwidth]{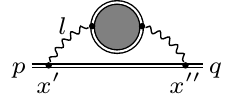}
	\caption{\label{fig:diagram} The bubble-chain electron mass operator, see Eq.~\eqref{m_def}. 
	}
\end{figure*}
Let us now consider the electron mass operator in a CCF with accounting for the polarization corrections to the virtual photon as depicted in Fig.~\ref{fig:diagram}. In the current work, our derivation is based on the Morozov-Ritus approach to calculation of the one-loop mass operator in a CCF \cite{morozov1975elastic} (and is different from the one adopted in Ref.~\cite{mironov2020resummation}). It relies on the specific properties of a CCF, which allow for significant simplifications.

The mass operator in the Ritus $E_p$-representation reads
\begin{align}
\label{m_def}
-i\Sigma(q,p)=&\Lambda^{2(D-4)}\int d^D x'\, d^D x''\,\bar{E}_{q}(x'')(ie\gamma^\mu) S_0^c(x'',x')(ie\gamma^\nu) E_{p}(x') D^c_{\mu\nu}(x'',x').
\end{align}
Here, the 4-momenta of the ingoing and outgoing electron are denoted by $p^\mu$ and $q^\mu$, respectively. In addition, by $l^\mu$ we denote the 4-momentum of the virtual photon (see Fig.~\ref{fig:diagram}). The integration is performed over the spacetime position of the vertices $x'$ and $x''$. 

We plug the $E_p$-functions \eqref{Ep}, the LO electron propagator $S_0^c$  \eqref{Sc} and the photon propagator $D^c_{\mu\nu}$ of a general form \eqref{Dx} into the expression \eqref{m_def}. We will calculate the spacetime integrals in Eq.~\eqref{m_def} in order to obtain $\Sigma(q,p)$ in the proper time representation. The intermediate computations are too lengthy to be presented in the paper, so instead we will outline key steps emphasizing some important details. All specific formulas, though, can be found in \cite{github}.

Let us start with expanding the integrand preexponential factor. Tedious $\gamma$-matrix algebra and tensor contractions is done with the aid of FeynCalc package \cite{mertig1991feyn,shtabovenko2016new,shtabovenko2020feyncalc}. The expression simplifies greatly if one applies the properties of a CCF, which can be done straightforwardly by substituting $F^{\mu\nu}=k^\mu a^\nu-k^\nu a^\mu$ and $(F^2)^{\mu\nu}=-a^2k^\mu k^\nu$ beforehand. An important observation at this stage is that the resulting expression does not contain terms with products of more than three $\gamma$-matrices. Furthermore, we rewrite the latter using the identity $\gamma^\alpha\gamma^\beta\gamma^\lambda=g^{\alpha\beta}\gamma^\lambda-g^{\alpha\lambda}\gamma^\beta+g^{\beta\lambda}\gamma^\alpha-i \varepsilon_{\delta \alpha\beta\lambda} \gamma^\delta\gamma^5$.

It is convenient to introduce new integration variables $x=x''-x'$, $X=(x'+x'')/2$ and the corresponding phase variables $\varphi=(kx)$, $\Phi=(kX)$. Then Eq.~\eqref{m_def} can be written in the form:
\begin{gather}
\Sigma(q,p)= - \frac{e^{-i \frac{\pi}{2}D}\alpha m}{(4\pi)^{D+1}}\int d^D x\, d^D X\, \int_0^\infty \frac{ds}{s^{D/2}}\, \int_0^\infty \frac{dt}{t^{D/2}}\, \Gamma\, e^{i \Theta},\\
\Gamma=\sum\limits_{n=0}^2 \mJ_n(t,\chi_l) \Gamma_n
\end{gather}
where we denoted the $\gamma$-matrix factor and the integrand phase by $\Gamma$ and $\Theta$, respectively. The functions $\mJ_n(t,\chi_l)$ are given by \eqref{J_def}. We will use these shorthand notations throughout the Section. At the current step, we have
\begin{eqnarray}
\label{Theta}
\begin{split}
\Theta=&-m^2s- \frac{x^2}{4\omega}+\frac12\left[(px)+(qx)\right]-(pX)+(qX)+ e(ax)\Phi\\
&+\frac{e(ap)}{8(kp)} (\varphi-2\Phi)^2-\frac{e(aq)}{8(kq)} (\varphi+2\Phi)^2+\frac{m^2\xi^2}{48}\left[ \frac{(\varphi-2\Phi)^3}{(kp)}+\frac{(\varphi+2\Phi)^3}{(kq)}-4s\varphi^2\right],
\end{split}
\end{eqnarray}
where we introduced the notation $\omega^{-1}=s^{-1}+t^{-1}$. At the same time, $\Gamma_n$ depends on $x'$, $x''$ only through $x$ and $\Phi$. 

Let us now pass to the light-cone variables: $x^\mu\rightarrow \lbrace x_-=\varphi/m,x_+,\vec{x}_\perp\rbrace$,  $X^\mu\rightarrow \lbrace X_-=\Phi/m,X_+,\vec{X}_\perp\rbrace$. Then one may see that $X_+$ and $\vec{X}_\perp$ enter $\Theta$ linearly, as $(pX)=p_-X_+ + p_+\Phi/m- \vec{p}_\perp\vec{X}_\perp$ [and the same for $(qX)$]. By integrating them out, we obtain the conservation of the minus and transverse components of the incoming electron momentum:
\[
\int d X_+ X^{D-2}_\perp e^{i(q_--p_-)X_+ - i (\vec{q}_\perp - \vec{p}_\perp)\vec{X}_\perp}\ldots \longrightarrow (2\pi)^{(D-1)}\delta(q_--p_-)\delta^{(D-2)}(\vec{q}_\perp - \vec{p}_\perp)\ldots .
\]
Now, noting that $(kq)=(kp)$ and $(aq)=(ap)$ allows one to simplify the expressions considerably. In particular, these substitutions cast $\Gamma_n$ into the form
\begin{equation}
\label{Gamma_n}
\Gamma_n=\mathcal{S}_n+\mathcal{V}^{(1)}_{n} (\gamma x) + \mathcal{V}^{(2)}_{n} e^2 (\gamma F^2 x)+\mathcal{V}^{(3)}_{n} e(\gamma F x) + \mathcal{T}_n e(\sigma F)+\mathcal{A}_n ie(\gamma F^\star x)\gamma^5,
\end{equation}
where for each of the  $\gamma$-matrix structures, we introduced a scalar factor, which depends only on the proper times $s$, $t$ and (polynomially) on the phases $\varphi$, $\Phi$. Let us note that the term $\propto (\gamma F)_{\mu}x^\mu$ should vanish in the final result due to the charge symmetry \cite{ritus1972radiative}. 

We proceed with rewriting the scalar products with $x$ in terms of light-cone variables, which yields:
\[
\Theta=\left[ (kp)-\frac{\varphi}{2\omega} \right]\frac{x_+}{m} + \ldots,\quad\Gamma=\Gamma'x_+ + \ldots,
\]
where we omitted terms independent of $x_+$. Noteworthy, in the expression \eqref{Gamma_n}, the $x_+$-terms originate only from $(\gamma x)=[(\gamma k)/m]x_+ + \ldots$. Hence, we may integrate out $x_+$ and $\varphi$ with the aid of the identity
\begin{equation}
\label{rel1}
\int_{-\infty}^\infty d\varphi\,f(\varphi) e^{ig(\varphi)}
\int_{-\infty}^\infty dx_+ \left[
\renewcommand*{\arraystretch}{0.8}
\begin{array}{c}
1 \\ x_+
\end{array} \right]
e^{i x_+ P(\varphi)} = 
\frac{2\pi  e^{ig(\varphi)}}{|P'_\varphi(\varphi)|}
\left.
\left[
\renewcommand*{\arraystretch}{0.8}
\begin{array}{c}
f(\varphi) \\ i\left( \frac{f(\varphi)}{P'_\varphi(\varphi)} \right)'_\varphi - f(\varphi)\frac{g'_\varphi(\varphi)}{P'_\varphi(\varphi)}
\end{array} \right]
\right|_{\varphi=\varphi_0},
\end{equation}
where  $\varphi_0$ is a zero of $P(\varphi)$, which in our case is $\varphi_0=2\omega (kp)$. So far we mainly followed Ref.~\cite{morozov1975elastic}. However, we proceed differently in what follows, as for the considered here photon propagator \eqref{Dx} the functions  $\mJ_n(t,\chi_l)$ inexplicitly depend on $\varphi$, see Eq.~\eqref{chi_l}. As a result, the integration in Eq.~\eqref{rel1} introduces into $\Gamma$ the terms that are proportional to $(\mJ_n)'_{\chi_l}$:
\[
\Gamma=\sum\limits_{n=0}^2 \left[\mJ_n(t,\chi_l) \Gamma_n+ (\mJ_n)'_{\chi_l} \tilde\Gamma_n \right].
\]
Note that $(\mJ_0)'_{\chi_l}=0$ since $D_0$ is independent of $\chi_l$ [see Eq.~\eqref{D_0}]. 

After the previous step, $\Theta$ can be represented as
\begin{equation}
\begin{split}
\Theta=\frac{\vec{x}_\perp^2}{4\omega}-(e\vec{a}_\perp \Phi+\vec{p}_\perp)\vec{x_\perp}+\ldots,
\end{split}
\end{equation} 
while $\Gamma$ is linear in $\vec{x}_\perp$. Hence, the $d^{D-2}x_\perp$ integral is Gaussian and can be calculated straightforwardly. After the integration, all the terms in $\Gamma$ that are dependent on $\Phi$ cancel out, so $\Phi$ enters the integrand only though the phase: $\Theta=(q_+-p_+)\Phi/m+\ldots$. The integral over $\Phi$ results into the electron's momentum $p_+$ component conservation. Therefore, the bubble-chain mass operator, given by Eq.~\eqref{m_def}, is diagonal in the $E_p$-representation:
\begin{equation}
\label{diag}
\Sigma(q,p)= \Lambda^{D-4} (2\pi)^D \delta^{(D)}(q-p) \Sigma(p,F).
\end{equation}
Hereinafter, we will consider only the diagonal part $\Sigma(p,F)$. 

We have integrated out all the spatial variables in Eq.~\eqref{m_def} but left the proper time integrals. It is convenient to restore the covariant notations now. $\Gamma$ is then spanned by the matrices $(\gamma p)$, $(\gamma F^2 p)$, $(\sigma F)$ and $(\gamma F^\star)\gamma^5$ (as expected, the matrix $\gamma^\mu F_{\mu\nu}$ is absent \cite{ritus1972radiative}). 

Let us introduce the following change of variables $\{s,t\}\rightarrow\{u,\sigma\}$:
\begin{equation}
\label{proper_times}
s=s(u,\sigma)=\frac{1+u}{m^2\chi^{2/3} u^{1/3}}\sigma,\quad t=t(u,\sigma)=\frac{s(u,\sigma)}{u},
\end{equation}
where $\chi=\xi (kp)/m^2$ is the electron dynamical parameter. Then the phase of the integrand reads
\begin{gather}
\label{Theta_sigma}
\Theta=-\frac{\sigma^3}{3}-z\sigma,\\
\label{z}
z=\left(\frac{u}{\chi}\right)^{2/3}\left[1-\frac{1}{u}\left(\frac{p^2}{m^2}-1\right)\right].
\end{gather} 

In the next step, we reexpress the factors $(\mJ_{1,2})'_{\chi_l}$ in $\Gamma$. The variable $u$ has the meaning of the dynamical parameter splitting ratio $\chi_l=u\chi/(1+u)$, hence, we may write $(\mJ_{1,2})'_{\chi_l}=\left[ (\mJ_{1,2})'_{u}-(\mJ_{1,2})'_{t} t'_u\right]/(\chi_l)'_u$. The term $(\mJ_{1,2})'_{u}$ can be integrated over $u$ by parts, taking into account that $\mJ_{1,2}(t,0)=0$ and vanishing of the integrand at $u\rightarrow\infty$. As for the second term, in effect, the derivative over $t$ replaces the functions $\mJ_{1,2}$ by $\tilde{\mJ}_{1,2}$ (up to a complex factor), see Eqs.~\eqref{J_def}, \eqref{Jt_def}. It is important to note that the last step is justified for $n=1,2$ as the corresponding terms are finite. 

To finalize the calculation, we integrate some terms in the resulting expression by parts over $\sigma$, specifically, the ones that are proportional to $\mJ_{1,2}(t(u,\sigma),\chi_l)e^2(\gamma F^2 p)\sigma^{-D/2}$. We rewrite them by applying the integral equality\footnote{We assume that the $\sigma$-integral are convergent as $\mJ_{1,2}\rightarrow0$ at $\sigma\rightarrow0$, see also \cite{mironov2020resummation}.}
\[
\int_0^\infty \frac{d\sigma}{\sigma^{D/2}} h(\sigma)e^{-i\sigma^3/3-iz\sigma}=\frac{2}{D-2}\int_0^\infty  \frac{d\sigma}{\sigma^{D/2-1}}\left[h'_\sigma(\sigma)-i(\sigma^2+z)h(\sigma)\right] e^{-i\sigma^3/3-iz\sigma}.
\]

Let us briefly discuss the renormalization procedure.\footnote{The full expression for the unrenormalized mass operator in $D$ dimensions is given in \cite{github}.} The divergent part of $\Sigma(p,F)$ reads
\begin{equation}
\begin{split}
\Sigma^0(p,F)=-\frac{ e^{-i\frac{\pi}{4} D}\alpha m^{D-3}\Lambda^{4-D}}{(4\pi)^{D/2-1}\chi^{(D-4)/3}} & \int_{0}^\infty \frac{du\, u^{2(D-4)/3}}{(1+u)^{D-2}}\\
&\times \int_{0}^\infty \frac{d\sigma}{\sigma^{D/2-1}}\left[ D- \frac{D-2}{1+u}\frac{(\gamma p)}{m}\right]e^{-i\sigma^3/3-iz\sigma}.
\end{split}
\end{equation}
The dimensional regularization allows applying various renormalization schemes \cite{coquereaux1980renormalization}. As the divergence is of the vacuum nature, we choose to subtract the field-free part $\Sigma(p,F=0)=\Sigma^0(p,F=0)$ and renormalize it on shell, i.e. $\Sigma(p,F)\longrightarrow \Sigma^{\text{o.s.}}_R(p,F=0)+[\Sigma(p,F)-\Sigma(p,F=0)]$, where the index $R$ stands for a renormalized quantity (see also \cite{ritus1972radiative,baier1976theory,meuren2015nonlinear}).  As the expression for $\Sigma(p,F)$ is now finite and regular, we set $D\rightarrow 4$.

Finally, the diagonal part of the renormalized bubble-chain mass operator can be represented in the following form:
\begin{equation}
\label{mass_op_final}
\begin{split}
\Sigma(p,F)=\sum\limits_{n=0}^2 &\left[ m S_n(p^2,\chi)+(\gamma p) V^{(1)}_{n}(p^2,\chi)+\frac{e^2(\gamma F^2 p)}{m^4\chi^2} V^{(2)}_{n}(p^2,\chi)\right.\\
&\left.\quad +\frac{e(\sigma F)}{m\chi} T_n(p^2,\chi)+ \frac{e(\gamma F^\star p)\gamma^5}{m^2\chi}A_n(p^2,\chi)\right],
\end{split}
\end{equation}
where each of the $\gamma$-matrix structures is multiplied by an invariant scalar function. For $n=0$, these functions read:
\begin{widetext}
\begin{subequations}
\begin{eqnarray}
\label{mass_op0_coeff_S}
S_0(p^2,\chi)=\frac{\alpha}{\pi}\int_0^\infty \frac{du}{(1+u)^2}\left\lbrace\frac{2u+1}{(1+u)u_\lambda}-\log\left[1-\frac{1}{u} \left(\frac{p^2}{m^2}-1\right)\right] +f_1(z)\right\rbrace,\\
V^{(1)}_{0}(p^2,\chi)=-\frac{\alpha}{2\pi}\int_0^\infty \frac{du}{(1+u)^3}\left\lbrace 2\frac{2u+1}{u_\lambda}-\log\left[1-\frac{1}{u} \left(\frac{p^2}{m^2}-1\right)\right] + f_1(z) \right\rbrace,\\
V^{(2)}_{0}(p^2,\chi)=\frac{\alpha}{3\pi}\int_0^\infty du \frac{3+u}{(1+u)^2}\left(\frac{\chi}{u}\right)^{2/3} f'(z),\\
\label{V2_0}
T_0(p^2,\chi)=-\frac{\alpha}{2\pi}\int_0^\infty \frac{du}{(1+u)^2}\left(\frac{\chi}{u}\right)^{1/3} f(z),\\
\label{mass_op0_coeff_A}
A_0(p^2,\chi)=\frac{\alpha}{2\pi}\int_0^\infty du\frac{2+u}{(1+u)^3} \left(\frac{\chi}{u}\right)^{1/3} f(z),
\end{eqnarray}
\end{subequations}
\end{widetext}
where $u_\lambda=u+(m_\gamma/m)^2(1+u)/u$ and $z$ given in Eq.~\eqref{z}. We also introduced a small photon mass $m_\gamma$ eliminating the IR divergence in the field-free part of the mass operator. In these expressions we used the Ritus functions \cite{ritus1972radiative}:
\[
\begin{split}
f(z)=i\int_0^\infty d\sigma e^{-i\sigma^3/3-iz\sigma},\\
f_1(z)=\int_0^\infty \frac{d\sigma}{\sigma+i0}\left( e^{-i\sigma^3/3-iz\sigma}-1\right).
\end{split}
\]
Eqs.~\eqref{mass_op0_coeff_S}-\eqref{mass_op0_coeff_A} correspond to the one-loop mass operator in a CCF. Note that we express in the form which is similar to the one used in Ref.~\cite{narozhny1979radiation}.

The $n=1,2$ terms in Eq.~\eqref{mass_op_final} give the nontrivial bubble-chain contribution. In particular, the corresponding scalar functions are given by
\begin{widetext}
\begin{subequations}
\begin{eqnarray}
\label{mass_op_coeff_S}
S_{1,2}(p^2,\chi)=\frac{i\alpha}{8\pi^2}\int_0^\infty \frac{du\, u}{(1+u)^2} \int_0^\infty \frac{d\sigma}{\sigma} \mJ_{1,2}\, e^{-i\frac{\sigma^3}{3}-iz\sigma},\\
\label{mass_op_coeff_V1}
V^{(1)}_{1,2}(p^2,\chi)=-\frac{i\alpha}{8\pi^2}\int_0^\infty \frac{du}{(1+u)^3} \int_0^\infty \frac{d\sigma}{\sigma} \mJ_{1,2}\, e^{-i\frac{\sigma^3}{3}-iz\sigma},\\
\label{mass_op_coeff_V2}
\begin{split}
V^{(2)}_{1,2}(p^2,\chi)=&\frac{i\alpha}{16\pi^2} \int_0^\infty \frac{du}{(1+u)^2} \int_0^\infty d\sigma  \left\lbrace 2\left(\frac{u^2+2u+2}{1+u}\pm 1\right)\left(\frac{\chi}{u}\right)^{2/3}\sigma \mJ_{1,2}\right.\\
&  +\left.\left[ \left(1+\frac{2}{u}-\frac{u^2+u+2}{u(1+u)}\frac{p^2}{m^2} \right)\mJ_{1,2} + \frac{u^2+2u+2}{u^2} \tilde{\mJ}_{1,2} \right]\frac{1}{\sigma} \right\rbrace e^{-i\frac{\sigma^3}{3}-iz\sigma},
\end{split}\\
\label{mass_op_coeff_T}
T_{1,2}(p^2,\chi)=\frac{\alpha}{16\pi^2} \int_0^\infty \frac{du}{1+u} \int_0^\infty d\sigma \left( \frac{1}{1+u}\pm 1 \right)\left(\frac{\chi}{u}\right)^{1/3} \mJ_{1,2}e^{-i\frac{\sigma^3}{3}-iz\sigma},\\
\label{mass_op_coeff_A}
A_{1,2}(p^2,\chi)=-\frac{\alpha}{8\pi^2} \int_0^\infty \frac{du}{(1+u)^2} \int_0^\infty d\sigma \left( \frac{1}{1+u}\pm 1 \right)\left(\frac{\chi}{u}\right)^{1/3} \mJ_{1,2}e^{-i\frac{\sigma^3}{3}-iz\sigma}.
\end{eqnarray}
\end{subequations}
\end{widetext}
In these expressions, $z$ is given in Eq.~\eqref{z}, and we imply that $\mJ_{1,2}=\mJ_{1,2}\left(t(u,\sigma), u\chi/(1+u)\right)$ (and the same for $\tilde{\mJ}_{1,2}$).

\subsection{Electron propagator}
\label{sec:exact_electron_propagator}
Radiative corrections cast Eq.~\eqref{S_eq} for the LO electron propagator into the DS equation. In the $E_p$-representation it takes a simple algebraic form:
\begin{equation}
\label{Dyson_el}
- i D(p,F)S^c(p,F)=-i\left[(\gamma p)-m-\Sigma(p,F)\right]S^c(p,F)=1,
\end{equation} 
where we used the diagonality of the mass operator. In a CCF, it is a general property of the matrix $D(p,F)$ that it can be decomposed into a sum of $\gamma$-matrix terms \cite{ritus1972radiative}:
\begin{equation}
D(p,F)=m S+(\gamma p) V^{(1)}+\frac{e^2(\gamma F^2 p)}{m^4\chi^2} V^{(2)} +\frac{e(\sigma F)}{m\chi} T+ \frac{e(\gamma F^\star p)\gamma^5}{m^2\chi}A,
\end{equation}
where the scalar coefficients $S$, $V^{(1,2)}$, $T$ and $A$ are functions of $p^2$ and $\chi$. By substituting Eq.~\eqref{mass_op_final} into $D(p,F)$, we find these coefficient in our case: 
\begin{equation}
\begin{split}
S=-1-\sum\limits_{n=0}^2 S_n,\,\, V^{(1)}=1-\sum\limits_{n=0}^2 V^{(1)}_n, \\
V^{(2)}=-\sum\limits_{n=0}^2 V^{(2)}_n,\,\, T=-\sum\limits_{n=0}^2 T_n,\,\,   A=-\sum\limits_{n=0}^2 A_n.
\end{split}
\end{equation}
Then it is possible to write out the electron propagator $S^c(p,F)=iD^{-1}(p,F)$ explicitly \cite{ritus1972radiative}:
\begin{widetext}
\begin{equation}
\label{Sc_exact}
\begin{split}
S^c(p,F)=i & \left[m S-(\gamma p) V^{(1)}-\frac{e^2(\gamma F^2 p)}{m^4\chi^2} V^{(2)}\right.\\
&\quad \left. -\frac{e(\sigma F)}{m\chi} T+ \frac{e(\gamma F^\star p)\gamma^5}{m^2\chi}A\right]\sum\limits_{\pm}\frac{1\pm (\gamma n)\gamma^5}{2 D_\pm},
\end{split}
\end{equation}
\end{widetext}
where $n_\mu=e(F^\star p)_\mu/m^3\chi$, and $D_\pm=D_\pm(p^2,\chi)$ factorize $\det D(p,F)=D_+D_-$ and read
\begin{equation}
D_\pm=m^2 S^2-p^2 V^{(1)\,2}+ m^2\left( A^2-2 V^{(1)}V^{(2)} \right)  \pm 2m^2 \left(S A -2 T V^{(1)}\right).
\end{equation}
At vanishing $\alpha$, the expression in Eq.~\eqref{Sc_exact} corresponds to Eq.~\eqref{Sc_Ep}. Note that $S^c(p,F)$ has an infinite number of poles corresponding to the solutions of the equations $D_\pm=0$. Note that an more detailed analysis of $S^c(p,F)$ properties can be found in Ref.~\cite{ritus1972radiative}.

By applying the $E_p$-transformation to Eq.~\eqref{Sc_exact}, we obtain the bubble-chain electron propagator in the proper time representation:
\begin{widetext}
\begin{equation}
\label{Sc_exact_x}
\begin{split}
S^c(x'',x')= &e^{i(ax)\Phi} e^{-i \frac{\pi}{2}\frac{D-4}{2}}  \frac{ \Lambda^{4-D}}{(4\pi)^{D/2+1}}\int_0^\infty\frac{ds}{s^{D/2}} \exp\left\lbrace-i\frac{x^2}{4s}+i\frac{s}{12}e^2\left(Fx\right)^2\right\rbrace
\\
&\times\left[m \mathcal{S}(s,\varphi)+\frac{(\gamma x)}{2s}\mathcal{V}^{(1)}(s,\varphi)+\frac{e^2(\gamma F^2 x)}{m^2\xi^2 \varphi^2}\mathcal{V}^{(2)}(s,\varphi) +\frac{e(\sigma F)}{m\xi\varphi}\mathcal{T}(s,\varphi) \right.\\
&\quad\quad \quad\quad\quad\quad\quad\quad\quad\quad\left.+\frac{e(\gamma F^\star x)\gamma^5}{\xi\varphi}\mathcal{A}(s,\varphi)+\frac{e(\gamma F^\star x)\gamma^5(\gamma x)}{2ms\xi\varphi}\mathcal{A}'(s,\varphi)\right].
\end{split}
\end{equation}
\end{widetext}
where the scalar coefficients $\mathcal{S}$, $\mathcal{V}^{(1,2)}$, $\mathcal{T}$, $\mathcal{A}$ and $\mathcal{A'}$ accommodate integrals over $p^2$ from the scalar functions utilized in Eq.~\eqref{Sc_exact}. The explicit expressions for them are lengthy and thus relegated to Appendix \ref{app1}. The physical meaning of these functions is smearing of the proper time causal step function $\theta(s)$ due to accounting for the radiative corrections to the LO propagator. In a sense, the scalar coefficients in Eq.~\eqref{Sc_exact_x} are analogous to the functions $\mJ_{1,2}$ in the bubble-chain photon propagator, see Eqs.~\eqref{J_def}, \eqref{Jt_def}. 

As the LO electron propagator [see Eq.~\eqref{Sc}], $S^c(x'',x')$ factorizes into the nondiagonal exponent $e^{i(ax)\Phi}$ and the diagonal part, which depends on $x'$, $x''$ only through $\varphi$. The $\gamma$-matrix structure of Eq.~\eqref{Sc_exact_x} differs from Eq.~\eqref{Sc} only by an additional term $\propto (\gamma F^\star x)\gamma^5(\gamma x)$ arising due to the spin factor $[1\pm (\gamma n)\gamma^5]$ in Eq.~\eqref{Sc_exact} (which reduces to 1 at the LO).

\subsection{Electron elastic scattering amplitude}
\label{sec:amplitude}
For completeness, let us present the on-shell electron elastic scattering amplitude $T_s(p)=-\mathcal{M}(\chi)/(2p^0)$, where $\mathcal{M}(\chi)\equiv \bar{u}_{p,\lambda} \Sigma(p,F)|_{p^2=m^2} u_{p,\lambda}$ is the invariant amplitude. It can be derived from Eq.~\eqref{mass_op_final} straightforwardly by applying the following equalities \cite{meuren2015nonlinear}: 
\begin{equation}
\label{amplitude_rules}
\bar{u}_p(\gamma p)u_p=2m,\quad \bar{u}_p \,e^2(\gamma F^2p)\,u_p= 2m^6\chi_p^2,\quad \bar{u}_p \,e (\sigma F) \, u_p= 4  s^\mu e F^*_{\mu\nu} p^\nu,
\end{equation}
where $s^\mu=(\bar{u}_p \gamma^\mu \gamma^5 u_p)/2m$ is the electron spin 4-vector \cite{berestetskii1982quantum}. Then the invariant amplitude reads:
\begin{equation}
\label{amplitude_general}
\begin{split}
\mathcal{M}(\chi)=\sum\limits_{n=0}^2 & \left\lbrace 2m^2\left[ S_n(m^2,\chi)+ V^{(1)}_{n}(m^2,\chi)+ V^{(2)}_{n}(m^2,\chi) \right] \vphantom{\frac{2 e(s F^\star p)}{m\chi}}\right.\\
& \quad \left. + \frac{2 e(s F^\star p)}{m\chi}\left[2T_n(m^2,\chi)+ A_n(m^2,\chi)\right]\right\rbrace.
\end{split}
\end{equation}
Whereas this result is of general form, $\mathcal{M}(\chi)$ in the bubble-chain approximation is obtained by the substitution of Eqs.~\eqref{mass_op0_coeff_S}-\eqref{mass_op_coeff_A}. The amplitude $\mathcal{M}$ naturally splits into $\mathcal{M}=\mathcal{M}^{(0)}+\delta\mathcal{M}$, where $\mathcal{M}^{(0)}$ corresponds to $n=0$ and $\delta\mathcal{M}$ contains the rest terms. Then $\mathcal{M}^{(0)}$ coincides with the one-loop scattering amplitude and reads
\begin{widetext}
	\begin{eqnarray}
	\label{M0_1}
	\begin{aligned}
	&\mathcal{M}^{(0)}(\chi)=\frac{\alpha m^2}{\pi} \int_{0}^{\infty} \frac{du}{(1+u)^2}\,  \\
	&\quad\quad\quad\quad \times \left[\frac{2u+1}{1+u}f_1(z_0) + \frac23 (u+3)\left(\frac{\chi}{u}\right)^{2/3}f'(z_0) -\frac{2\gamma_s}{1+u} \left(\frac{u}{\chi}\right)^{2/3} f(z_0)  \right].
	\end{aligned}
	\end{eqnarray}
\end{widetext}
Here, we introduced $z_0=(u/\chi)^{2/3}$ and $\gamma_s=e F^*_{\mu\nu}s^\mu p^\nu/2m^3$.\footnote{Note that in Ref.~\cite{mironov2020resummation} $\gamma_s$ is introduced with an opposite sign, which is though compensated by the definition of $\gamma^5$ (and hence $s^\mu$), that also differs by sign from the one adopted in this paper.}  The identity
\begin{eqnarray}
\int_{0}^{\infty} \frac{du}{(1+u)^2} \left[ \frac{2(u-2)}{3(1+u)}\left(\frac{\chi}{u}\right)^{2/3}f'(z_0)-\frac{2u}{1+u} f_1(z_0) \right]=0
\end{eqnarray}
casts Eq.~\eqref{M0_1} into the more familiar form \cite{ritus1972radiative}:
\begin{widetext}
	\begin{eqnarray}
	\begin{aligned}
	\label{M0}
	&\mathcal{M}^{(0)}(\chi)=\frac{\alpha m^2}{\pi} \int_{0}^{\infty} \frac{du}{(1+u)^2}\,  \\
	&\quad\quad\quad\quad \times \left[f_1(z_0) +  \frac{u^2+2u+2}{1+u}\left(\frac{\chi}{u}\right)^{2/3}f'(z_0) -\frac{2\gamma_s}{1+u} \left(\frac{u}{\chi}\right)^{2/3} f(z_0)  \right],
	\end{aligned}
	\end{eqnarray}
\end{widetext}
The other term $\delta\mathcal{M}=\delta\mathcal{M}_1+\delta\mathcal{M}_2$ corresponds to the nontrivial contribution from the 1PI bubble-chain corrections and is given by:
\begin{widetext}
	\begin{equation}
	\label{deltaM}
	\begin{split}
	\delta{\cal M}_{1,2}(\chi)=&\frac{i\alpha m^2}{(2\pi)^2}\int_{0}^{\infty}\frac{du}{(1+u)^2}\int_0^\infty d\sigma e^{-i\sigma^3/3-iz_0 \sigma} \left\lbrace \left(\frac{u^2+2u+2}{1+u}\pm 1\right)\left(\frac{\chi}{u}\right)^{2/3}\sigma \mJ_{1,2} \vphantom{\left(\frac{u}{\chi}\right)^{2/3}}\right. \\
	&\left.+\left[ \mJ_{1,2} + \frac{u^2+2u+2}{2 u^2} \tilde{\mJ}_{1,2}\right]\frac1\sigma  -2i\gamma_s\left(\frac{1}{1+u}\pm 1 \right) \left(\frac{u}{\chi}\right)^{2/3} \mJ_{1,2} \right\rbrace  ,
	\end{split}
	\end{equation}
\end{widetext}
where, as previously, $\mJ_{1,2}=\mJ_{1,2}\left(t(u,\sigma), u\chi/(1+u)\right)$ (and the same for $\tilde{\mJ}_{1,2}$). Note that Eq.~\eqref{deltaM} is equivalent to Eq.~(16) from Ref.~\cite{mironov2020resummation}.

\section{Discussion}
\label{sec:discussion}
The high-$\chi$ scaling of the mass operator $\Sigma(p,F)$, the electron propagator $S^c(p,F)$ and the scattering amplitude $\mathcal{M}(\chi)$ [given in Eqs.~\eqref{mass_op_final}, \eqref{Sc_exact} and \eqref{amplitude_general}] is determined by the asymptotic properties of the scalar functions $S_n$, $V^{(1,2)}_n$, $T_n$ and $A_n$ [see Eqs.~\eqref{mass_op0_coeff_S}-\eqref{mass_op_coeff_A}]. The asymptotic properties of the one-loop contribution, given by Eqs.~\eqref{mass_op0_coeff_S}-\eqref{mass_op0_coeff_A}, was studied by Narozhny in Ref.~\cite{narozhny1979radiation}. He identified the terms dominating at $\chi\gg 1$, namely $V^{(1,2)}_0$, while the rest can be omitted. Here, using similar argumentation, we study the nontrivial part represented by Eqs.~\eqref{mass_op_coeff_S}-\eqref{mass_op_coeff_A}.

We start with comparing $T_{1,2}$, $A_{1,2}$ [Eqs.~\eqref{mass_op_coeff_T}, \eqref{mass_op_coeff_A}] with $V^{(2)}_{1,2}$; in particular, with the term that is proportional to $\sigma\mJ_{1,2}$ [Eq.~\eqref{mass_op_coeff_V2}]. The integral formation scales in these expressions are of the same order. At the same time, the $\sigma\mJ_{1,2}$ term in $V^{(2)}_{1,2}$ is enhanced by the factor of $\chi^{2/3}$, while $T_{1,2}$ and $A_{1,2}$ contain only $\chi^{1/3}$. Therefore, we will neglect the latter two terms against $V^{(2)}_{1,2}$.

The functions $S_{1,2}$ and $V^{(1)}_{1,2}$ share similar integral structure [see Eqs.~\eqref{mass_op_coeff_S}, \eqref{mass_op_coeff_V1}]. Let us estimate the latter. To this end, we represent $\mJ_{1,2}$ as in Eq.~\eqref{J_res}, so that $V^{(1)}_{1,2}$ reads:
\begin{equation}
\label{sigma_int1}
V^{(1)}_{1,2}\sim \alpha\int_0^\infty \frac{du}{(1+u)^3} \int_0^\infty \frac{d\sigma}{\sigma} \left[ e^{-i\Pi_{1,2}(0,\chi_l)t(u,\sigma)}-1\right] e^{-i\sigma^3/3-iz\sigma},
\end{equation}
where $\chi_l=u\chi/(1+u)$ is the photon dynamical parameter (denoted as in Ref.~\cite{mironov2020resummation}), and we omitted numerical constants. The integrals are formed at $u\sim u_{\mathrm{eff}}\sim 1$ and $\sigma<\sigma_{\mathrm{eff}}\sim 1$. Hence, $t(u,\sigma)$ and $z$, given by Eqs.~\eqref{proper_times} and \eqref{z}, can be estimated as $t\sim \sigma/m^2\chi^{2/3}$ and $z\sim -\nu/m^2\chi^{2/3}$. Here, by $\nu=p^2-m^2$ we denote the virtuality of the incoming electron. Assuming that $\chi\gg 1$ and $\Pi_{1,2}(0,\chi)\ll m^2\chi^{2/3}$, we expand the exponent inside the brackets of Eq.~\eqref{sigma_int1} up to the linear term, which gives:
\begin{equation}
\label{sigma_int2}
V^{(1)}_{1,2}\sim \frac{\alpha \Pi_{1,2}(0,\chi)}{{m^2\chi^{2/3}}}  \int_0^1 d\sigma e^{-i\sigma^3/3+i\sigma \nu/m^2\chi^{2/3}}.
\end{equation}
Note that here $\chi_l\sim (\chi_l)_{\mathrm{eff}}\sim\chi$.\footnote{The effective scales in $V^{(1)}_{1,2}$ are the same as in the term $\delta \mathcal{M}^{(I)}$ in Ref.~\cite{mironov2020resummation}, see Table II therein.} $S_{1,2}$ can be estimated similarly.

Let us now consider the two cases of small and large values of $\nu/m^2\chi^{2/3}$, assuming $\chi\gg 1$. At small virtuality $\nu\ll m^2\chi^{2/3}$, including the on-shell case, we get the estimate $S_{1,2}\sim V^{(1)}_{1,2}\sim \alpha \Pi_{1,2}(0,\chi)/m^2\chi^{2/3}\ll \alpha$. It means that $S_{1,2}$ and $V^{(1)}_{1,2}$ are negligible in comparison to $V^{(2)}_{1,2}$, which is enhanced by $\chi^{2/3}$. We, therefore, conclude that the dominating terms in the amplitude $\mathcal{M}(\chi\gg 1)$ are $V^{(2)}_{n}$. It is noteworthy that the two asymptotic contributions to $\mathcal{M}(\chi)$ considered in Ref.~\cite{mironov2020resummation} originate from this term.

In the opposite case, $\nu\gg m^2\chi^{2/3}$, we have $S_{1,2}\sim V^{(1)}_{1,2}\sim \alpha \Pi_{1,2}(0,\chi)/\nu$. While $S_{1,2}$ can be omitted straightforwardly, we have to be more careful with the functions $V^{(1)}_{1,2}$. In the mass operator $\Sigma(p,F)$ and the electron propagator $S^c(p,F)$, they are multiplied by $(\gamma p)\sim \nu$, hence, $\nu V^{(1)}_{1,2}\sim \alpha \Pi_{1,2}(0,\chi)$. Let us compare it with the term proportional to $\tilde\mJ_{1,2}/\sigma$ in $V^{(2)}_{1,2}$ [see Eq.~\eqref{mass_op_coeff_V2}]. To the same accuracy as in Eq.~\eqref{sigma_int1}, we estimate this term as
\begin{equation}
\tilde V^{(2)}_{1,2}\sim \frac{\alpha}{m^2}\int_0^\infty du\frac{u^2+2u+2}{u^2} \int_{\sigma_0(u)}^\infty \frac{d\sigma}{\sigma} \Pi_{1,2}(0,\chi_l) e^{-i\Pi_{1,2}(0,\chi_l)t(u,\sigma)} e^{-i\sigma^3/3-iz\sigma},
\end{equation}
where we used Eq.~\eqref{Jt_res} and the fact that $\tilde\mJ_{1,2}$ effectively cuts off the $\sigma$-integral from below \cite{mironov2020resummation}. As in Eq.~\eqref{sigma_int1}, the integrand is proportional to $\Pi_{1,2}(0,\chi_l)$, but in contrary the $u$-integral is formed at $u\sim u_{\mathrm{eff}}\sim 1/\chi$  \cite{mironov2020resummation}. The latter means that $\tilde V^{(2)}_{1,2}$ is enhanced by a factor of $\chi$, hence, larger than $\nu V^{(1)}_{1,2}$. 

\begin{figure}
	\begin{tabular}{cc}
		\includegraphics[valign=c,width=0.49\linewidth]{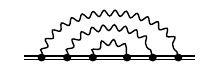} & \includegraphics[valign=c,width=0.49\linewidth]{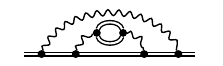} \\
		(a) & (b)
	\end{tabular}\\
	\begin{tabular}{c}
	\includegraphics[valign=c,width=0.4\linewidth]{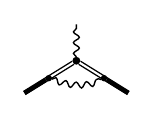}\\
	(c)
	\end{tabular}
	\caption{Three-loop corrections to the electron propagator (a,b); a one-loop correction to the bare vertex connecting two exact fermion propagators (c).}
	\label{fig:other_corrections}
\end{figure}

We conclude that in the nonperturbative regime $\alpha\chi^{2/3}>1$, the asymptotically dominant contribution to $\Sigma(p,F)$, $S^c(p,F)$ and $\mathcal{M}(\chi)$ is given by the scalar function $V^{(2)}_{n}(p^2,\chi)$. Let us emphasize that such a situation is specific to the bubble-chain approximation. For loop corrections of other types, the term $(\gamma p)V^{(1)}_{n}$ might also become relevant. For instance, in the asymptotic studies of the two- and three-loop contributions \cite{narozhny1979radiation,narozhny1980expansion}, Narozhny observed that the $V^{(1)}_{0}$ term in $\Sigma(p,F)$ should be accounted for in the rainbow-type corrections, obtained by successive insertions of the one-loop mass operator as shown in Fig.~\ref{fig:other_corrections}(a). Notably, such corrections are enhanced only logarithmically (see also Ref.~\cite{mironov2020resummation} for a more detailed review). At the same time, for the insertions of the two-loop bubble-type mass operator as in Fig.~\ref{fig:other_corrections}(c), the asymptotic behaviour of the total correction was determined by $V^{(2)}_{n}$ in agreement with our findings. 

According to the asymptotic analysis of the one-loop vertex function $\Gamma^\mu$ \cite{di2020one,morozov1981fian}, the dominating scaling $g=\alpha\chi^{2/3}$ arises from the terms $\propto (\gamma k) k^\mu $. However, such terms do not contribute to scattering amplitudes due to gauge invariance \cite{di2020one}. Let us now consider a one-loop correction to a vertex that connects two exact electron propagators as shown in Fig.~\ref{fig:other_corrections}(c). Note that the mass correction to the vertex is already taken into account in the exact propagators. The correction will be enhanced by a factor of $g$ if the product of dominating terms in the exact propagator and $\Gamma^\mu$ survives. However, the asymptotically strongest contribution to the electron propagator is given by the term $\propto(\gamma F^2 p)V^{(2)}=-a^2(\gamma k)(kp)V^{(2)}$. Hence, when the leading terms are multiplied, they vanish as $(\gamma k)^2=0$. Therefore, the vertex correction should be enhanced by a factor that is asymptotically weaker than $g$. This supposition is in favour of the RN conjecture, but should be substantiated with a full length calculation to be presented elsewhere.

Finally, let us discuss the implications of the presented results for the DS equations. The overall tensor and $\gamma$-matrix structure of the photon and electron propagators in a CCF given in Eqs.~\eqref{photon_prop}, \eqref{Dx} and \eqref{Sc_exact}, \eqref{Sc_exact_x}, respectively, is of general form. It also applies to the exact propagators, i.e. a self-consistent solution of Eqs. \eqref{Dyson_ph} and \eqref{Dyson_el} (see also Fig.~\ref{fig:DS}). In order to obtain the proper scalar functions pertaining to such a solution, one should recalculate the polarization and mass operators with account for the electron propagator $S^c$ given by Eq.~\eqref{Sc_exact}. When it is rewritten in the proper time representation $S^c(x'',x')$  as in Eq.~\eqref{Sc_exact_x}, its structure appears to be not much different from that of $S^c_0(x'',x')$ [Eq.~\eqref{Sc}]. Therefore, we suppose that the calculation of the exact mass operator is feasible and should generally follow the steps outlined in Sec.~\ref{sec:mass_op}. The first part of this supposition also applies to the exact photon polarization operator. In effect, such a self-consistent computation to be presented elsewhere will enclose the system of the DS equations and reformulate them in terms of the scalar coefficients $\Pi_{1,2}$, $S$, $V^{(1,2)}$, $T$ and $A$. Furthermore, the resulting system can be simplified at $g>1$ by taking into account only the dominating contributions  in the bubble-chain approximation.

\section{Conclusions}
\label{sec:conclusions}
We have calculated the electron mass operator in a CCF that combines the LO electron propagator and the bubble-chain photon propagator of general form. The latter accounts for the resummed Dyson series of 1PI polarization loop insertions. To simplify the computations, we have determined the proper time representation for the propagators. The resulting expression for the mass operator is expanded over 5 $\gamma$-matrix structures multiplied by scalar invariant functions of $p^2$ and $\chi$, which we present explicitly. This expansion is of general form, therefore allows to determine the corresponding bubble-chain electron propagator and the elastic scattering amplitude.

The asymptotic properties of the electron mass operator, propagator and scattering amplitude are determined by the scalar coefficient functions $S_n$, $V^{(1,2)}_n$, $T_n$ and $A_n$. In the bubble-chain approximation, they are given by Eqs.~\eqref{mass_op0_coeff_S}-\eqref{mass_op_coeff_A}, and the dominating contribution at $\chi\gg 1$ is provided by $V^{(2)}_n$. This confirms the previous findings from the asymptotic studies at the 3-loop level \cite{narozhny1979radiation,narozhny1980expansion} and of the bubble-chain scattering amplitude \cite{mironov2020resummation}. The dominance of this particular term might lead to effective suppression of the vertex corrections.

The discussed here photon and electron propagator structures are general for a CCF and apply to the DS equations. The presented here calculations can be generalized by replacing the leading order electron propagator with the bubble-chain one. This will further allow formulating the DS equations in the bubble-chain approximation in terms of the scalar functions. Furthermore, we hope that our computer-algebraic scripts \cite{github} will make such a lengthy computation feasible.

\section*{Acknowledgements}
A.A.M. was supported by the Russian Foundation for Basic Research (Grant No. 19-32-60084). 
A.M.F. was supported by the MEPhI Academic Excellence Project (Contract No. 02.a03.21.0005), Russian Foundation for Basic Research (Grants No. 19-02-00643 and No. 20-52-12046).

\appendix
\section{Exact electron propagator in the proper time representation}
\label{app1}
The exact electron propagator in the coordinate representation can be calculated from Eq.~\eqref{Sc_exact} by applying the transformation 
\begin{equation}
\label{Sc_Ep1}
S^c(x'',x')=\Lambda^{4-D} \int \frac{d^Dp}{(2\pi)^D}E_p(x'')S^c(p,F)\bar{E}_p(x'),
\end{equation}
Let us outline the main steps of the evaluation (for a detailed derivation see \cite{github}). We start with expanding the gamma matrix prefactor in the integrand and change the variables $x'$, $x''$ to $x=x''-x'$, $X=(x'+x'')/2$, as we did above. Then we express all the scalar products with $p$ in terms of the light-cone variables $p_{\pm,\perp}$, and perform the substitutions $p_+=(p^2+\vec{p}_\perp^2)/2p_-$, $p_-=x_-/2s$. Next, we pass to the new integration variables $\lbrace s,\,p^2,\,\vec{p}_\perp^2 \rbrace$, so that $d^Dp=(1/|2s|)ds\,dp^2\,d^{D-2}p_\perp$. We proceed by evaluating the Gaussian integral over $d^{D-2}p_\perp$ and return to covariant notation for scalar products. Note that after this step, the terms proportional to $p^2$ enter the preexponential factor. We remove these terms using integration by parts in $s$. Finally, by collecting the $\gamma$-matrix structures, we arrive to Eq.~\eqref{Sc_exact_x}, where the coefficients can be expressed as follows:
\begin{widetext}
\begin{subequations}
	\begin{eqnarray}
	\label{Sc_exact_coeff_S}
	\mathcal{S}(s,\varphi)=\sum\limits_{\zeta=\pm}\int_{-\infty}^\infty\frac{dp^2\,e^{-ip^2 s}}{D_\zeta(p^2,\chi)}\left\lbrace S(p^2,\chi)-\zeta\frac{i}{2} \xi \varphi V^{(1)}(p^2,\chi)+\zeta A(p^2,\chi)\right\rbrace,\\
	\mathcal{V}^{(1)}(s,\varphi)=-\sum\limits_{\zeta=\pm}\int_{-\infty}^\infty\frac{dp^2\,e^{-ip^2 s}}{D_\zeta(p^2,\chi)}V^{(1)}(p^2,\chi),\\
	\begin{split}
	\mathcal{V}^{(2)}(s,\varphi)=m^2\sum\limits_{\zeta=\pm}\int_{-\infty}^\infty & \frac{dp^2\,e^{-ip^2 s}}{D_\zeta(p^2,\chi)}\left\lbrace i \zeta s  \xi\varphi S(p^2,\chi) +\frac{s}{3}\xi^2\varphi^2 V^{(1)}(p^2,\chi)  \vphantom {\left[\frac{V^{(1)}}{D_\zeta}\right]'_\chi}\right. \\  
	&-2sV^{(2)}(p^2,\chi) -4 \zeta s T(p^2,\chi) + is \xi\varphi A(p^2,\chi) \\
	& \quad\quad\quad\quad\quad \left. -\frac{i \xi\varphi}{2m^4 s} D_\zeta(p^2,\chi) \left[\frac{V^{(1)}(p^2,\chi)}{D_\zeta(p^2,\chi)}\right]'_\chi \right\rbrace,
	\end{split}\\
	\begin{split}
	\mathcal{T}(s,\varphi)= m^2& \sum\limits_{\zeta=\pm} \zeta \int_{-\infty}^\infty  \frac{dp^2\,e^{-ip^2 s}}{D_\zeta(p^2,\chi)}\left\lbrace i \zeta s  \xi\varphi S(p^2,\chi)   \vphantom {\left[\frac{V^{(1)}}{D_\zeta}\right]'_\chi}\right. \\
	&+\left(\frac{s}{3}\xi^2\varphi^2-i\frac{D-3}{m^2}\right) V^{(1)}(p^2,\chi)-2sV^{(2)}(p^2,\chi) -4 \zeta s T(p^2,\chi)  \\
	& \quad\quad\quad\quad\quad\quad \left. + is \xi\varphi A(p^2,\chi) -\frac{i \xi\varphi}{2m^4 s} D_\zeta(p^2,\chi) \left[\frac{V^{(1)}(p^2,\chi)}{D_\zeta(p^2,\chi)}\right]'_\chi \right\rbrace,
	\end{split}\\
	\mathcal{A}(s,\varphi)= \sum\limits_{\zeta=\pm} \zeta \int_{-\infty}^\infty\frac{dp^2\,e^{-ip^2 s}}{D_\zeta(p^2,\chi)}\left\lbrace S(p^2,\chi)-\zeta\frac{i}{2} \xi \varphi V^{(1)}(p^2,\chi)+\zeta A(p^2,\chi)\right\rbrace,\\
	\label{Sc_exact_coeff_A1}
	\mathcal{A'}(s,\varphi)=- \sum\limits_{\zeta=\pm} \zeta \int_{-\infty}^\infty\frac{dp^2\,e^{-ip^2 s}}{D_\zeta(p^2,\chi)} V^{(1)}(p^2,\chi),
	\end{eqnarray}
\end{subequations}
\end{widetext}
where $\chi=\xi(kp)/m^2= \xi\varphi/2 m^2 s$.


%

\end{document}